\documentclass{aastex631}

\usepackage[whole]{bxcjkjatype}

\begin{document}

\title{An ALMA Glimpse of Dense Molecular Filaments Associated with High-mass Protostellar Systems in the Large Magellanic Cloud}

\author[0000-0002-2062-1600]{Kazuki Tokuda}
\affiliation{Department of Earth and Planetary Sciences, Faculty of Science, Kyushu University, Nishi-ku, Fukuoka 819-0395, Japan}
\affiliation{National Astronomical Observatory of Japan, National Institutes of Natural Sciences, 2-21-1 Osawa, Mitaka, Tokyo 181-8588, Japan}

\author[0000-0002-8217-7509]{Naoto Harada}
\affiliation{Department of Earth and Planetary Sciences, Faculty of Science, Kyushu University, Nishi-ku, Fukuoka 819-0395, Japan}

\author[0000-0002-6907-0926]{Kei E. I. Tanaka}
\affiliation{Department of Earth and Planetary Sciences, Tokyo Institute of Technology, Meguro, Tokyo, 152-8551, Japan}

\author[0000-0002-7935-8771]{Tsuyoshi Inoue}
\affiliation{Department of Physics, Konan University, Okamoto 8-9-1, Kobe, Japan}

\author[0000-0002-0095-3624]{Takashi Shimonishi}
\affiliation{Environmental Science Program, Faculty of Science, Niigata University, Ikarashi-ninocho 8050, Nishi-ku, Niigata, 950-2181, Japan}

\author[0000-0001-7511-0034]{Yichen Zhang}
\affiliation{Department of Astronomy, University of Virginia, Charlottesville, VA 22904-4325, USA}
\affiliation{RIKEN Cluster for Pioneering Research, Wako, Saitama 351-0198, Japan }

\author[0000-0003-2248-6032]{Marta Sewi{\l}o}
\affiliation{Exoplanets and Stellar Astrophysics Laboratory, NASA Goddard Space Flight Center, Greenbelt, MD 20771, USA}
\affiliation{Department of Astronomy, University of Maryland, College Park, MD 20742, USA}
\affiliation{Center for Research and Exploration in Space Science and Technology, NASA Goddard Space Flight Center, Greenbelt, MD 20771}

\author[0000-0002-9627-1600]{Yuri Kunitoshi}
\affiliation{Department of Physics, Graduate School of Science, Osaka Metropolitan University, 1-1 Gakuen-cho, Naka-ku, Sakai, Osaka 599-8531, Japan}

\author[0000-0002-4098-8100]{Ayu Konishi}
\affiliation{Department of Physics, Graduate School of Science, Osaka Metropolitan University, 1-1 Gakuen-cho, Naka-ku, Sakai, Osaka 599-8531, Japan}

\author[0000-0002-8966-9856]{Yasuo Fukui}
\affiliation{Department of Physics, Nagoya University, Furo-cho, Chikusa-ku, Nagoya 464-8601, Japan}

\author[0000-0001-7813-0380]{Akiko Kawamura}
\affiliation{National Astronomical Observatory of Japan, National Institutes of Natural Sciences, 2-21-1 Osawa, Mitaka, Tokyo 181-8588, Japan}

\author[0000-0001-7826-3837]{Toshikazu Onishi}
\affiliation{Department of Physics, Graduate School of Science, Osaka Metropolitan University, 1-1 Gakuen-cho, Naka-ku, Sakai, Osaka 599-8531, Japan}

\author[0000-0002-0963-0872]{Masahiro N. Machida}
\affiliation{Department of Earth and Planetary Sciences, Faculty of Science, Kyushu University, Nishi-ku, Fukuoka 819-0395, Japan}

\begin{abstract}

Recent millimeter/sub-millimeter facilities have revealed the physical properties of filamentary molecular clouds in relation to high-mass star formation.
A uniform survey of the nearest, face-on star-forming galaxy, the Large Magellanic Cloud (LMC), complements the Galactic knowledge. We present ALMA survey data with a spatial resolution of $\sim$0.1\,pc in the 0.87\,mm continuum and HCO$^{+}$(4--3) emission toward 30 protostellar objects with luminosities of 10$^4$--10$^{5.5}$\,$L_{\odot}$ in the LMC. The spatial distributions of the HCO$^{+}$(4--3) line and thermal dust emission are well correlated, indicating that the line effectively traces dense, filamentary gas with an H$_2$ volume density of $\gtrsim$10$^5$\,cm$^{-3}$ and a line mass of $\sim$10$^3$--10$^{4}$\,$M_{\odot}$\,pc$^{-1}$. Furthermore, we obtain an increase in the velocity linewidths of filamentary clouds, which follows a power-law dependence on their H$_2$ column densities with an exponent of $\sim$0.5. This trend is consistent with observations toward filamentary clouds in nearby star-forming regions withiin $ \lesssim$1\,kpc from us and suggests enhanced internal turbulence within the filaments owing to surrounding gas accretion. Among the 30 sources, we find that 14 are associated with hub-filamentary structures, and these complex structures predominantly appear in protostellar luminosities exceeding $\sim$5 $\times$10$^4$\,$L_{\odot}$. The hub-filament systems tend to appear in the latest stages of their natal cloud evolution, often linked to prominent H$\;${\sc ii} regions and numerous stellar clusters. Our preliminary statistics suggest that the massive filaments accompanied by hub-type complex features may be a necessary intermediate product in forming extremely luminous high-mass stellar systems capable of ultimately dispersing the parent cloud.

\end{abstract}


\section{Introduction} \label{sec:intro}

Understanding the formation mechanism of high-mass stars ($>$8\,$M_{\odot}$) remains a compelling challenge in astrophysics due to their significant impact on the surrounding environment, both during their active accretion phase and beyond, into their main sequence stage. Over the past few decades, extensive theoretical and observational studies have been conducted to shed light on this energetic phenomenon (see reviews by e.g., \citealt{McKee_2007}, \citealt{Zinnecker_Yorke_2007}, and \citealt{Krumholz_2019}). The current understanding, both theoretical and observational, suggests that the formation of high-mass stars begins with the collapse of a massive ($\gtrsim$100\,$M_{\odot}$) dense core/clump, analogous to the formation of low-mass stars \cite[e.g,][]{McKeeTan_2003}. This gravitational collapse eventually leads to the formation of massive protostellar systems and the subsequent mass ejection through molecular outflows driven by the magnetic centrifugal wind from the circumstellar disks \cite[e.g.,][]{Machida_2013,Commercon_2022}, appears to be a common phenomenon across a wide range of stellar masses \citep{Beuther_2002,Maud_2015,Matsushita_2018}. Recent studies investigating distant, low-metallicity star-forming regions have provided further insights, indicating that the properties of molecular outflows are universal in diverse environments within the Local Group of galaxies, consistent with the observations and theories developed in our own Milky Way (\citealt{Tokuda_2022b} and references therein).

Despite significant theoretical and observational progress in understanding the evolution of individual star-forming units, i.e., dense pre- and protostellar cores, they do not exist in isolation. They are inextricably linked to the formation of larger gaseous structures, making it difficult to understand them from their formation onwards. Molecular clouds are hierarchically inhomogeneous, and their kinematics are largely affected by multiple physical processes, such as non-thermal turbulent motions and magnetic field (see the review by \citealt{Heyer_2015}). Observational characterization has advanced our understanding of their complex nature. In the solar neighborhood, molecular line studies using ground-based single-dish telescopes illustrated the quasi-universality of filamentary structures across low- and high-mass star-forming regions \cite[e.g.,][]{Mizuno_1995,Onishi_1996,Onishi_1998,Nagahama_1998} and the subsequent high-resolution dust continuum measurements with the Herschel Space Observatory revealed the detailed structures with a typical width of $\sim$0.1\,pc \citep[e.g.,][]{Arzounamian_2011,Arzoumanian_2019,Andre_2022}.
Some star-forming molecular clouds have a more complex shape with multiple filaments combined together and are sometimes referred to as $``$hub filament$"$ \citep{Myers_2009}. Their junction often becomes sites of active star formation \cite[e.g.,][]{Peretto_2013,Williams_2018}, and statistical studies compiling a large number of targets across the Galactic Plane suggest that they could be precursors evolving into stellar clusters \citep{Kumar_2020}.

High-resolution observations outside the Galaxy, especially in the Local Group members, provide complementary insights into the Galaxy thanks to the less line-of-sight contamination and uniform sampling catalogs, such as (high-mass) young stellar objects (YSOs), obtained in various wavelengths. Recent Atacama Large Millimeter/submillimeter Array (ALMA) CO observations toward some Local Group giant molecular clouds (GMCs) revealed a relatively diffuse molecular layer around dense proto-cluster clumps, and the formation of the above-mentioned complex hub-filaments is found to be associated with galaxy-scale converging gas flows \citep{Tokuda_2019,Tokuda_2020,Tokuda_2022a,Fukui_2019,Muraoka_2020}. 
However, previous studies of the Local Group have been inadequate in the following respects. (1) ALMA target selections have mostly been biased toward molecular clouds with bright infrared sources (see also Table~\ref{tab:target}), and the extreme star-forming environment, 30~Dor \citep{Indebeouw_2020,Wong_2022}. We still lack a general understanding of protocluster formation via (hub-type) filamentary clouds. (2) Molecular line tracers in high-density regions directly related to proto-cluster formation have not been thoroughly investigated.
From a more distant galactic perspective beyond the Local Group, the determination of star formation rates using high-density tracers such as HCN and HCO$^{+}$ has garnered much attention \citep[e.g.,][]{Gao_2004}, and has motivated further investigation of Galactic molecular clouds \citep[e.g.,][]{Shimajiri_2018, Torii_2019}. Small-scale, comprehensive studies using dense gas tracers in the Local Group may also provide clues to connect with large-scale phenomena occurring within galaxies, such as the evolution of GMCs \cite[e.g.,][]{Kawamura_2009,Chevance_2022}.

In this paper, we present ALMA observations toward 30 YSOs in the Large Magellanic Cloud (LMC) obtained as part of the MAGellanic Outflow and chemistry Survey (MAGOS) (PI: K. Tanaka, \#2019.1.01770.S). 
We utilize the 0.87\,mm continuum and HCO$^+$~(4--3) line data from the MAGOS survey to study the molecular cloud's evolution leading to high-mass proto-cluster formation, bridging a gap between the existing Galactic and extragalactic (the Local Group) studies. 
In Section~\ref{sec:obs}, we describe the observations and data reduction details. In Section 3, we investigate the role of HCO$^{+}$ as a high-density gas tracer by comparing the 0.87\,mm continuum. We performed the filamentary cloud identification in Section~\ref{R:filament_id}. We provide the discussion and implications of our analysis in Section~\ref{sec:dis}, and we summarize the overall results in Section~\ref{sec:summary}.

\section{Target Selection, ALMA Observations, and Data Reduction Details} \label{sec:obs}

The MAGOS project (PI: K. Tanaka, \#2019.1.01770.S) is a comprehensive study aimed at exploring various aspects of star formation within the Magellanic Clouds. The main target lines are CO~(3--2) as an outflow tracer and several hot-core tracer lines (e.g., multiple CH$_3$OH and SO$_2$ transitions). The specific data presented in this paper, which are part of this larger project, are used to investigate the properties of filamentary molecular clouds in the LMC. The observations were carried out using the ALMA 12\,m array in the C43-3 configuration, with the Band~7 receiver operating at 0.87\,mm. Our targets were selected from the massive YSO catalog with a total number of 277 in the LMC \citep{Seale_2009}. The YSOs were verified through spectroscopic observations with Spitzer's infrared spectrometer (5.3--38\,$\mu$m). Regarding the bolometric luminosity ($L_{\rm bol}$), we calculated them by directly integrating their Spectral Energy Distributions (SEDs) from 1$\mu$m to 500$\mu$m. The photometric data were taken from Two Micron All Sky Survey (2MASS), Spitzer, and Herschel archival data, after positional matching with the \cite{Seale_2009} YSO catalog. The derivation of $L_{\rm bol}$ is simply the integral of the wavelength data points, 1.215\,$\mu$m, 1.654\,$\mu$m, 2.157\,$\mu$m, 3.6\,$\mu$m, 4.5\,$\mu$m, 5.8\,$\mu$m, 8.0\,$\mu$m, 24\,$\mu$m, 100\,$\mu$m, 160\,$\mu$m, 250\,$\mu$m, 350\,$\mu$m, and 500\,$\mu$m without any SED modeling schemes. While the apparent luminosity can vary due to some protostellar nature, such as disk inclination angle, estimating the intrinsic luminosity is challenging. Nonetheless, we can achieve accuracy on a factor of $\sim$3 \citep{Zhang_2018}. 

We allocated a luminosity selection criteria of $>$10$^{4}$\,$L_{\odot}$ with young spectral features, such as polycyclic aromatic hydrocarbon emission, deep silicate absorption, fine-structure lines, and ice absorption. From a total number of 178 objects in the \cite{Seale_2009} catalog that fulfills the above criterion, we selected total 30 sources divided into three luminosity ranges: 10$^4$--10$^{4.5}$\,$L_{\odot}$ (Ll01,...,10), 10$^{4.5}$−10$^5$\,$L_{\odot}$ (Lm01,...,10), and 10$^5$--10$^{5.5}$\,$L_{\odot}$ (Lh01,...,10). This diverse luminosity-based target selection was motivated by our original observational purpose: to investigate how the nature of the CO outflows changes depending on their protostellar luminosities. Most of the targets were randomly selected 10 targets for each luminosity category, but some sources were arbitrarily selected based on the signs of molecular outflow or hot cores (a more detailed description can be found in Tanaka et al., in prep). In this paper, we have assigned new designations to these protostars to facilitate immediate discernment of their luminosities from their names. Hereafter, we refer to both the protostars and their associated molecular gas revealed with ALMA using the same names listed in Table~\ref{tab:target}.
Table~\ref{tab:target} provides information on target YSOs and associated GMCs (including evolutionary stages) and H$\;${\sc ii} regions \citep{Henize_1956,Fukui_2008,Kawamura_2009}. 
Multiple molecular lines are detected among the selected sources, with HCO$^+$~(4--3) being the second most intense line after CO~(3--2). This paper focuses on analyzing the HCO$^+$~(4--3) line and the 0.87\,mm continuum data.

\begin{table}[htb!]
\centering
\caption{Observation targets}
\label{tab:target}
\begin{tabular}{cccccccc}
\hline \hline
Name & \multicolumn{2}{c}{Coordinates\,(J2000)$^{\rm A}$} & $L_{\rm bol}$$^{\rm B}$ & Fukui08ID$^{\rm C}$ & GMCtype$^{\rm D}$ & HenName$^{\rm E}$ & Other name/reference$^{\rm F}$\\
& R.A.\,(deg) & Dec.\,(deg) & $(10^4 \,L_\odot)$ & & & & \\
\hline
Lh10 & 81.69421 & -68.81311 & 23.10 & 139 & III & N144B & ST11 (1) \\
Lh09 & 72.97204 & -69.39128 & 20.80 &  14 & III &  N79B & (2)\\
Lh08 & 84.91137 & -69.65119 & 16.50 & 191 & III & N160A & - \\
Lh07 & 78.35454 & -69.37919 & 15.50 &  72 & III & N113A & Region~A (3) \\
Lh06 & 77.46054 & -68.88486 & 15.30 &  63 & III & N105A & N105-1 (4)\\
Lh05 & 77.46775 & -68.89092 & 14.70 &  63 & III & N105A & N105-2 (4,5) \\
Lh04 & 84.92454 & -69.77000 & 12.70 & 197 & III &  N159 & N159W-South (6,7)\\
Lh03 & 79.80112 & -69.15203 & 11.60 &  98 &  II &  N119 & ST16 (8) \\
Lh02 & 84.90667 & -69.75694 & 11.30 & 197 & III &  N159 & N159W-North MMS-1 (6,9) \\
Lh01 & 84.90433 & -69.76019 & 10.30 & 197 & III &  N159 & N159W-North MMS-2 (6,9)\\
Lm10 & 72.79746 & -69.44631 &  8.30 & 234 &  II & N79   & - \\
Lm09 & 73.03842 & -66.92275 &  7.96 &  12 & III & N4B   & - \\
Lm08 & 82.72600 & -68.57453 &  7.76 & 154 & III & N148F & ST5 (10) \\
Lm07 & 84.66537 & -69.09394 &  6.18 & 186 & III & N157A & - \\
Lm06 & 84.87996 & -70.20469 &  4.58 & 197 & III & N171A & ST1 (10) \\
Lm05 & 78.32975 & -69.36417 &  4.21 &  72 & III & N113A & (2) \\
Lm04 & 80.51133 & -67.78392 &  3.68 &   - &   - &  N44J & - \\
Lm03 & 80.70783 & -66.68225 &  3.53 & 119 &  II &  N45A & - \\
Lm02 & 84.98192 & -71.16692 &  3.52 & 195 &  II & N214A & - \\
Lm01 & 73.96092 & -66.57628 &  3.25 &  23 & III & N11H  & - \\
Ll10 & 80.88917 & -69.62006 &  2.82 & 129 &  II & N132G & - \\
Ll09 & 80.50879 & -67.96489 &  2.81 & 114 & III &  N44C & - \\
Ll08 & 77.60037 & -70.23514 &  2.80 &   - &   - &  N110 & ST17 (10) \\
Ll07 & 81.59058 & -68.66642 &  1.93 & 137 &  II & N144B & - \\
Ll06 & 76.15871 & -70.91197 &  1.90 &  49 & III & N191A & - \\
Ll05 & 74.67696 & -66.14325 &  1.48 &  35 &  II &  N12  & - \\
Ll04 & 75.98387 & -67.34411 &  1.37 &  45 & III &  N17A & - \\
Ll03 & 84.47842 & -69.57661 &  1.13 & 184 &  II &  N158 & - \\
Ll02 & 73.53450 & -66.77539 &  1.12 &  19 &  II &  N6   & - \\
Ll01 & 81.50500 & -67.50336 &  1.08 & 251 &  II &  N51D & - \\
\hline
\end{tabular}\\
Table notes: [A] Spitzer source coordinate \citep{Seale_2009} [B] See the text in Sect.~\ref{sec:obs}. [C] Associated GMC catalogs with a $\sim$40\,pc resolution survey by \cite{Fukui_2008} [D] GMC evolutionary type classification by \cite{Kawamura_2009}: Type~II means GMCs with H$\;${\sc ii} regions, and Type~III represents GMCs with H$\;${\sc ii} regions and young massive clusters (see also Section~\ref{D:GMCevo}). [E] H$\;${\sc ii} region catalog by \cite{Henize_1956} [F] Previously observed in molecular line and/or millimeter/submillimeter continuum with ALMA 
(1) \cite{Shimonishi_2016}
(2) \cite{Nayak_2019} 
(3) \cite{Sewilo_2018} 
(4) \cite{Sewilo_2022} 
(5) \cite{Sewilo_2022HDO} 
(6) \cite{Fukui_2015} 
(7) \cite{Tokuda_2019} 
(8) \cite{Shimonishi_2020}
(9) \cite{Tokuda_2022a} 
(10) \cite{Shimonishi_2016VLT}
\end{table}

Using Common Astronomy Software Application (CASA; \citealt{CASA_2022}), we performed the imaging process based on the pipeline-calibrated visibility data provided by the observatory. For the process of the 0.87\,mm continuum images, we utilized two spectral windows covering frequencies of 343.9--344.9\,GHz and 345.8--347.7\,GHz.
We employed the \texttt{tclean} algorithm with the \texttt{multi-scale} deconvolver in the analysis. The imaging processes were carried out for the 0.87\,mm continuum and HCO$^+$ data, using the Natural and Briggs weighting with a robust parameter of 0.5, respectively. The resulting properties of the synthesized beams and the sensitivity are listed in Table~\ref{tab:data_prop}. Note that the observations were carried out using the same execution blocks for all targets, minimizing differences in observing time and elevation angle. This ensured consistent beam shapes and sensitivity across all observed fields.

\begin{table}[htb!]
\caption{Data properties}
\label{tab:data_prop}
\begin{tabular}{lccccc}
\hline
Line/Continuum         & Beam size                 & Beam P.A. & Velocity grid      & Aggregate bandwidth & R.M.S. sensitivity \\
\hline
\hline
0.87\,mm               & $\sim$0\farcs46 $\times$ 0\farcs39 & 30 deg.   & $\cdots$           & 2.8 GHz             & $\sim$0.2\,mJy\,beam$^{-1}$     \\
HCO$^{+}$ ($J$ = 4--3) & $\sim$0\farcs39 $\times$ 0\farcs32 & 25 deg.   & 0.5\,km\,s$^{-1}$ & $\cdots$           & $\sim$0.86\,K\\
\hline
\end{tabular}
\end{table}

\section{Results} \label{sec:res}

\subsection{Spatial distribtuions of 0.87\,mm continuum and HCO$^+$(4--3) emission} \label{R:HCO_cont}

\begin{figure}[htb!]
    \centering
    \includegraphics[width=0.9\columnwidth]{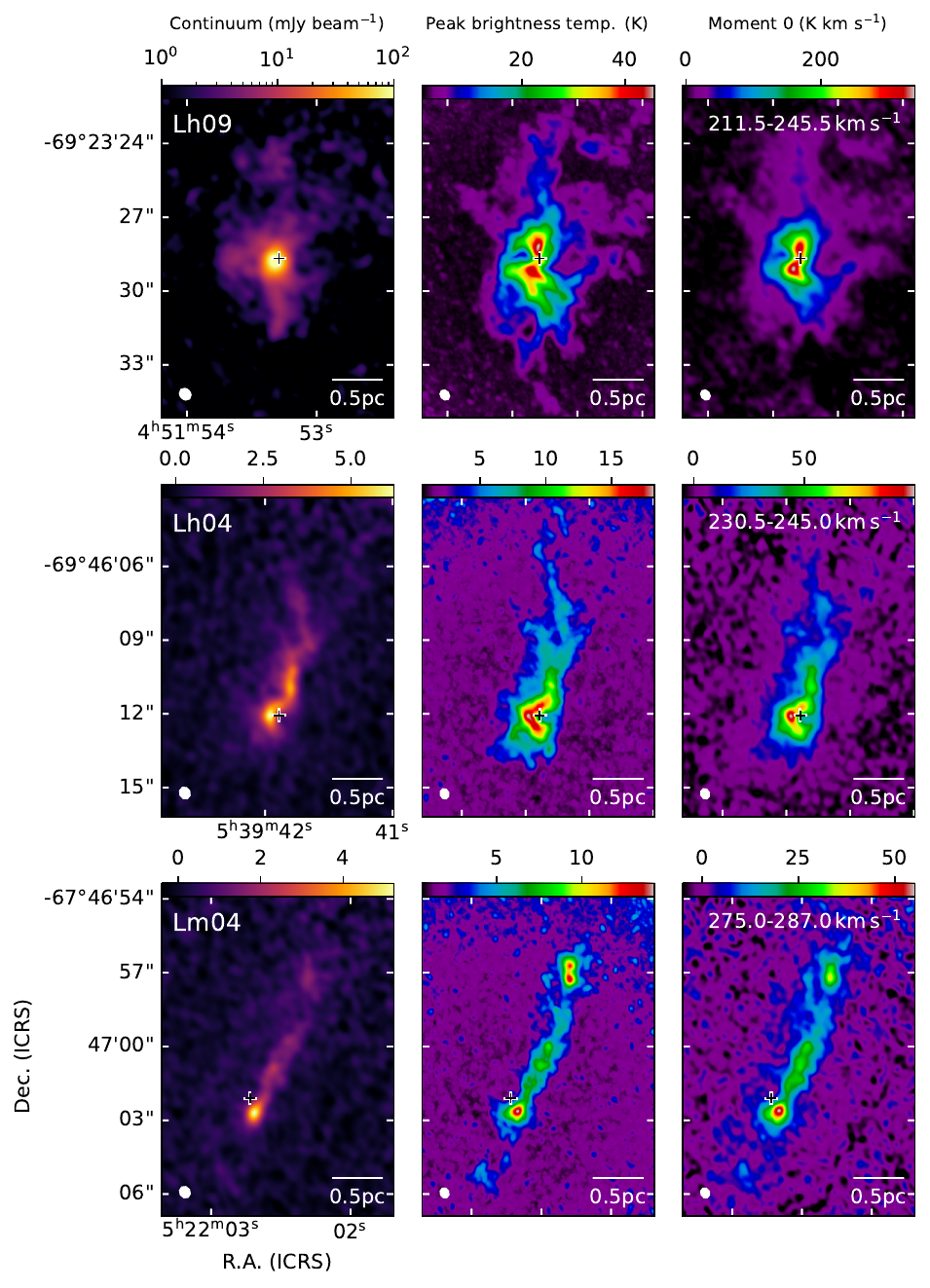}
    \caption{Enlarged views of the 0.87\,mm and HCO$^+$~(4--3) emission toward three example targets. The color-scale images show the distribution of the 0.87\,mm continuum emission (left panel), the HCO$^+$ peak brightness temperature (middle panel), and the velocity-integrated intensity (right panel). The crosses indicate the positions of massive protostars identified by Spitzer \citep{Seale_2009}. The velocity range used to derive the peak brightness temperature and moment~0 maps are indicated on the upper side of each right image. The synthesized beam size is indicated in white in the lower-left corner of each image.}
    \label{fig:HCO_cont}
\end{figure}

In this section, we discuss the spatial distribution of the 0.87\,mm continuum and HCO$^+$ emission. Figure~\ref{fig:HCO_cont} shows three examples in enlarged views close to the Spitzer-identified high-mass protostars. 
We chose these three because two of them (Lh09 and Lh04) have already been well studied by ALMA's early observations \citep{Nayak_2019,Tokuda_2019} with millimeter continuum and other molecular lines at different wavelengths, and their prominent filamentary structures can be immediately determined visually in the present data set, and their similar structures were confirmed associated with a less bright protostar for Lm04. Although the ALMA continuum peaks and Spitzer-identified protostar positions do not necessarily coincide exactly (Figure~\ref{fig:HCO_cont} left panels), possibly due to the multiwavelength point-source positional determination using the Spitzer data, the discrepancy is about an ALMA's single-beam element, $\sim$0\farcs4, which is within the accuracy of the Spitzer/IRAC survey's angular resolution, $\sim$2$\arcsec$ \citep{Meixner_2006}. 
Since there is only a very small spatial discrepancy, and the Spitzer sources have been spectroscopically identified as massive protostars (see Sect.~\ref{sec:obs}), it is safe to assume that the 0.87\,mm sources are indeed dense envelope gas associated with the protostars, rather than a coincidental line-of-sight overlap.
The brightest 0.87\,mm source in our sample, Lh09, shows the complicated structure. The less bright target, lm04 shows a simpler structure composed of a single-filamentary distribution. The 0.87\,mm continuum emission most likely traces the thermal dust emission of dense material around the high-mass protostellar sources. 

To characterize the physical properties of the dense molecular clouds traced by the 0.87\,mm continuum emission as a first step, we derived the total mass gas based on the 0.87\,mm flux density measurements. The flux density above $\sim$3$\sigma$ detection of Lh09, Lh04, and Lm04 are 505\,mJy, 102\,mJy, and 52\,mJy, respectively, which can be converted into gas mass by adopting an absorption coefficient per unit dust mass at 0.87\,mm, a dust-to-gas mass ratio and the dust temperature of 1.8\,cm$^2$\,g$^{-1}$, 3.5 $\times$10$^{-3}$, and 20\,K, respectively, with an optically thin case \citep[e.g.,][]{Ossenkopf_1994,Herrera_2013,Gordon_2014} in the LMC region of 50\,kpc \citep{Pietrzynski_2019}. The resultant total gas mass of Lh09, Lh04, and Lm04 is $\sim$2$\times$10$^{4}$\,$M_{\odot}$, $\sim$4$\times$10$^{3}$\,$M_{\odot}$ and $\sim$2$\times$10$^{3}$\,$M_{\odot}$, respectively. It should be noted that the expected uncertainties are a factor of two or three, which are more likely due to systematic errors related to dust properties and temperature than measurement errors.

The middle and right panels in Figure~\ref{fig:HCO_cont} show the peak brightness temperature and velocity-integrated intensity (moment0) maps of HCO$^{+}$~(4--3), respectively. Overall, the distributions of the 0.87\,mm continuum and HCO$^{+}$ are quite similar, suggesting that the emitting regions for both tracers are spatially correlated. For optically thin cases with a kinematic temperature of 10--100\,K, the critical density of HCO$^{+}$~(4--3) is (2--3) $\times$10$^{6}$\,cm$^{-3}$ \citep{Shirley_2015}, which is comparable to the average density derived from the continuum emission (see also Sect.~\ref{R:filament_id}). The observed intensities of HCO$^{+}$ exceeding 10\,K indicate a high likelihood of thermalization. For instance, in the case of Lh04, assuming the local thermo-dynamical equilibrium (LTE) conditions \citep{Rohlfs_2004} with a uniform excitation temperature of 20\,K througout the emitting region, the derived column density of HCO$^{+}$ is $\sim$3 $\times$10$^{13}$\,cm$^{-2}$.
Applying a literature value of HCO$^{+}$ abundance relative to H$_2$ in the LMC-N159 region \citep{Johansson_1994}, [HCO$^{+}$]/[H$_2$] = 1.8 $\times$10$^{-10}$, the H$_2$ column density is estimated to be 2 $\times$10$^{23}$\,cm$^{-2}$, which is consistent with those derived from the 0.87\,mm data (see Sect.~\ref{R:filament_id}). 
We confirmed that the integrated intensity of HCO$^{+}$ and 0.87\,mm flux density within a 1\,pc radius aperture show a tight, linear correlation with each other for all sources because Pearson's correlation coefficient and $p$-value are 0.963 and 2e-17, respectively.
Thus, the HCO$^{+}$(4--3) line is useful for tracing the highly dense regions of molecular clouds in the LMC. This effectiveness of HCO$^{+}$ as a high-density gas tracer has also been demonstrated in studies utilizing single-dish observations in the LMC \citep{Paron_2016,Galametz_2020} and other galaxies such as M33 \citep{Braine_2017}, which has a similar metallicity to the LMC.

\begin{figure}[htb!]
    \centering
    \includegraphics[width=1.0\columnwidth]{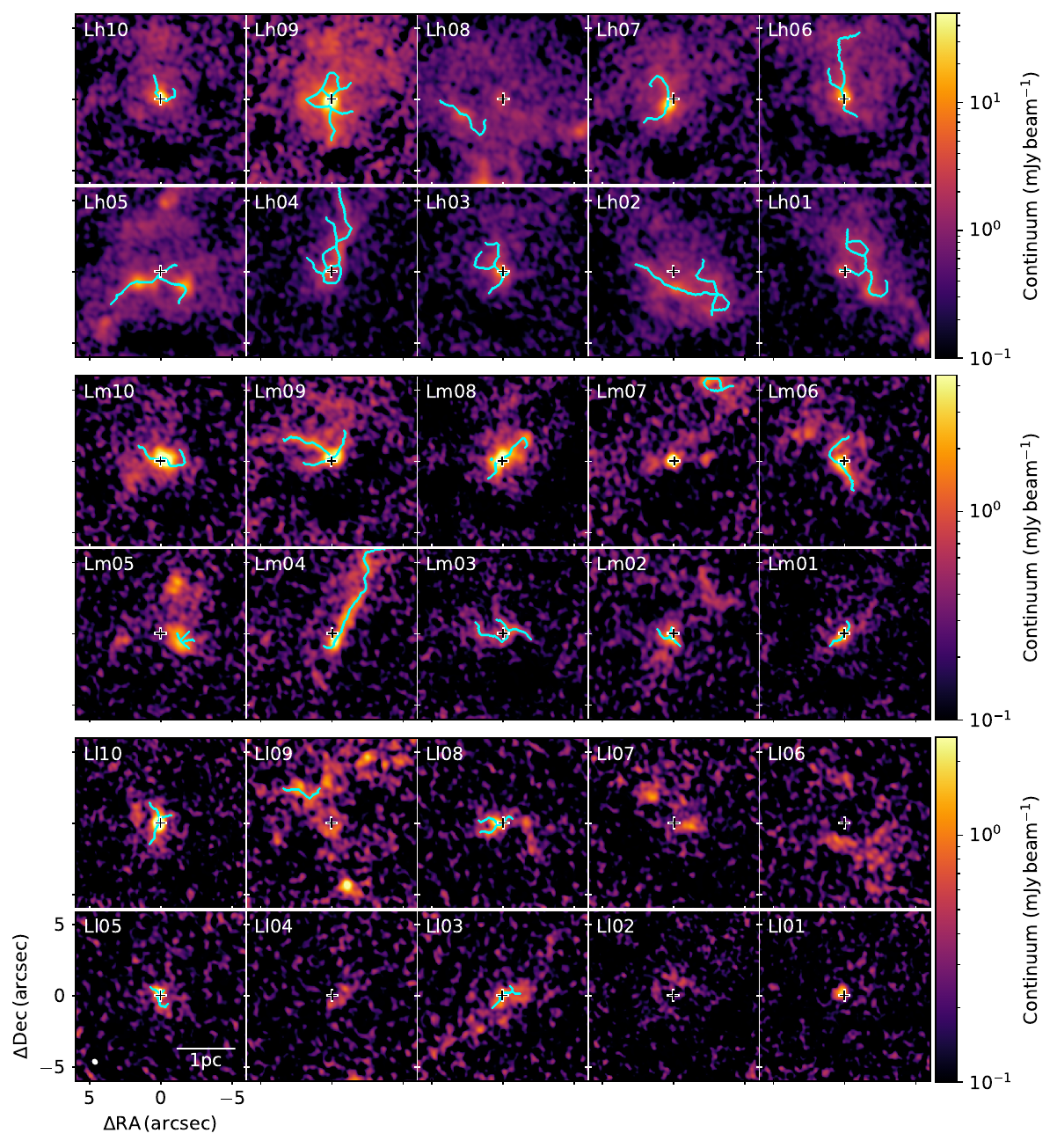}
    \caption{0.87\,mm continuum distributions toward high-mass YSOs in the LMC. The angular resolution, $\sim$0\farcs46 $\times$ 0\farcs39, is indicated with the white ellipse in the lower left corner of the lower left panel. Crosses represent the position of the YSOs \citep{Seale_2009}. 
    Cyan lines show the identified spines of the filament that we used to derive the physical properties (see the text in Sect.~\ref{R:filament_id}).
    }
    \label{fig:cont_view}
\end{figure}

Figure~\ref{fig:cont_view} presents the 0.87\,mm continuum emission for all sources. 
Our sample (Table~\ref{tab:target}) with 0.1\,pc-scale continuum observations has the largest number of targets among any other literature in the extragalactic study.
The spatial distributions are overall in agreement with previous studies (see the references in Table~\ref{tab:data_prop}) and show peaks either coinciding with or close to the positions of the YSOs. 
However, some sources have positional discrepancies larger than the ALMA beam size between the continuum peaks and YSO positions. 
For instance, while the continuum peak of Lh08 is significantly distanced from the position of the YSO, the structure through 0.87\,mm and HCO$^+$ shows that they are interconnected in both space and velocity (see Figures~\ref{fig:HCO_cont}, \ref{fig:HCO_view}, \ref{fig:MOM0_view} and \ref{fig:MOM1_view}). 
In the case of Lh05, according to \cite{Sewilo_2022}, the two sources were identified as one during the multiband-merging process of Spitzer. 
For Lh02, the presence of free-free emission could explain the offset between the position of the YSO and continuum (see YSO-N1 and the 1.3\,mm continuum source, MMMS-1 in \citealt{Tokuda_2022a}), but for the 0.87mm emission, the spatial correlation with dense gas tracers, C$^{18}$O (Figure~6 in \citealt{Tokuda_2022a}) and HCO$^+$ are quite good, suggesting that the 0.87\,mm continuum structure is predominantly capturing thermal emission. For Lh01, outflows have been reported for each continuum peak (see MMS-2a/b and MMMS-d in Figure~5 in \citealt{Tokuda_2022a}), indicating that the system is harboring multiple protostars within a few times 0.1\,pc. 
According to the Australia Telescope Compact Array (ATCA) observations at wavelengths of 3 and 6\,cm with a beam size of 1$\arcsec$--2$\arcsec$ by \cite{Indebetouw_2004}, in addition to the above-mentioned Lh09 and Lh02, Lh08 and Lh06 might also have some free-free contamination as their centimeter continuum peaks are close to those of 0.87\,mm. For example, in Lh09, \cite{Nayak_2019} reported the H30$\alpha$ detection at the continuum peak, indicating significant free-free contamination from ionized gas, but the emitting region is compact with a size scale of 0.1\,pc (Figure~2 in \citealt{Nayak_2019}). In such a case, the ionized contamination in 0.87\,mm does not extend to the surrounding filament gas. As shown here, some sources require careful handling but do not significantly affect the identification of the overall filament structure and a detailed discussion is given in Sect.~\ref{R:filament_id} and \ref{sec:dis}.

\subsection{Filamentary cloud identification} \label{R:filament_id}

\begin{figure}[htbp]
    \centering
    \includegraphics[width=1.0\columnwidth]{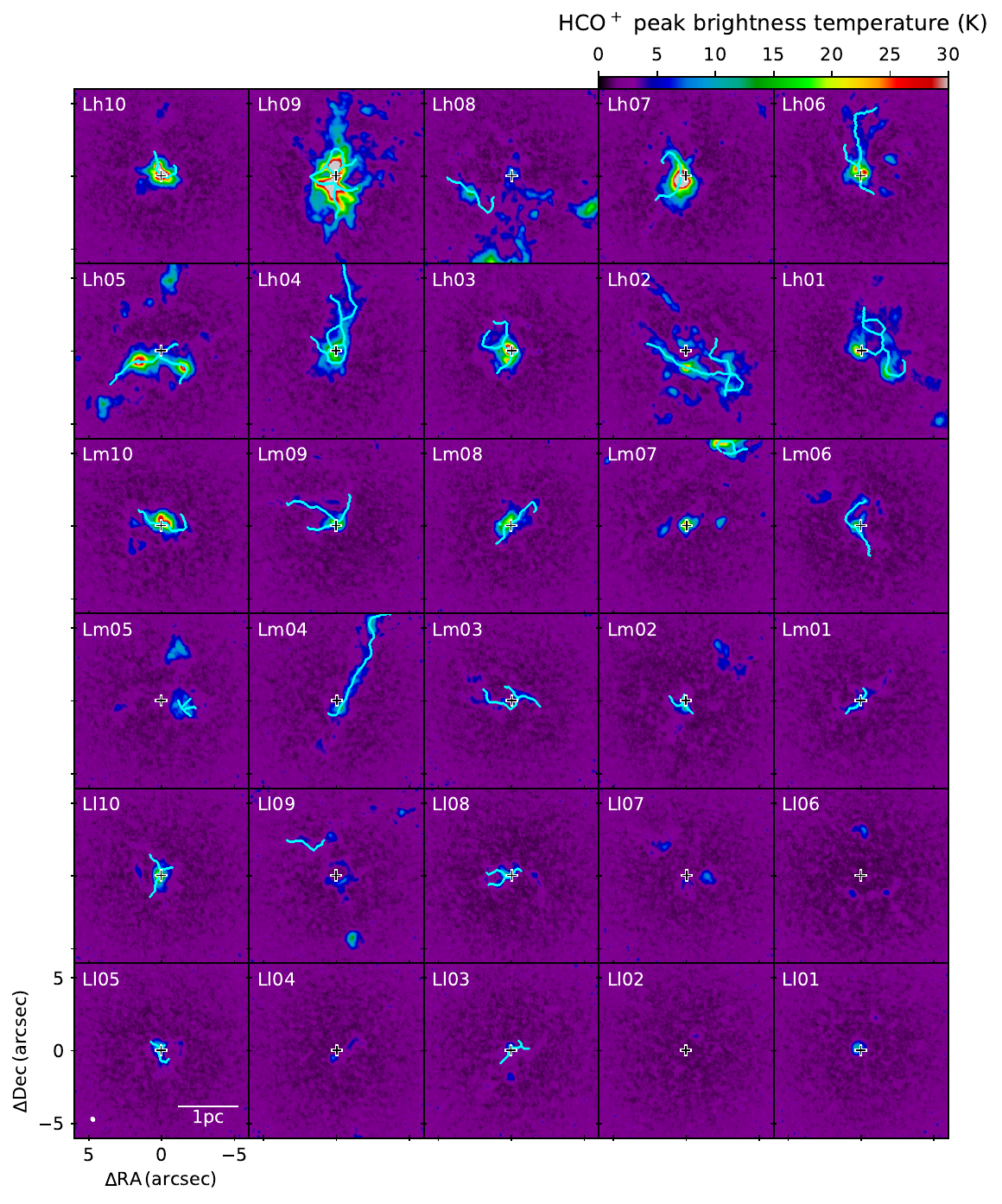}
    \caption{The distribution of the HCO$^+$~(4--3) peak brightness temperature toward high-mass YSOs in the LMC. Symbols and lines are the same as in Figure~\ref{fig:cont_view}.}
    \label{fig:HCO_view}
\end{figure}

As shown in Figures~\ref{fig:cont_view} and \ref{fig:HCO_view}, most of the targets have an elongated morphology. We used the Filfinder package \citep{Koch_2015} to extract such elongated structures based on certain criteria. 
We emphasize that the main purpose of this analysis is to characterize the physical quantities of primary elongated structures with uniform treatment. We applied the algorithm to the 0.87\,mm continuum images. 
In the filament identification analysis, we excluded areas with coarser sensitivity and high noise levels at the edges of the ALMA mosaics. For the used regions, the sensitivity is at least 50\% of the field's maximum, corresponding to a circular area with a radius of 8\farcs7 (= 2.1\,pc at 50\,kpc) from the center. In the initial step, prior to creating a masked image, we selected \texttt{flatten\_percent} = 95 to flatten the intensity distribution of sources with high contrast. Then, emission masks were created by applying a threshold of 3$\sigma$ noise level and selecting regions with an area greater than five times the beam size. The other input parameters were \texttt{adapt\_thresh} = 0.1\,pc and \texttt{smooth\_size} = 0.05\,pc, which are recommended values to extract filamentary structure whose width is close to $\sim$0.1\,pc \citep{Koch_2015}. The intensity distribution within the mask was used to identify the filament spines. Filaments shorter than five times the beam size and filament branches shorter than three times the beam size were excluded from the analysis.

The filament identification results are shown in Figures~\ref{fig:cont_view} and \ref{fig:HCO_view}. Although the algorithm identified more than two spatially separated filament spines for some fields, we illustrate the longest, nearest neighbor of protostar positions to characterize the primary filamentary cloud for each object. A global trend throughout the targets is that filamentary structures toward brighter sources are more extended and more complex: multiple filaments appear to intersect at the position of YSOs. 
For sources with luminosities below a luminosity $\sim$10$^{4}$\,$L_{\odot}$, HCO$^+$ emission is mostly detected only in a small region ($\lesssim$1\,pc) around the protostar (Figure~\ref{fig:HCO_view}), which means no surrounding dense filamentary layer in the fainter sources. 

We derived the physical quantities along the filament spine shown in the cyan lines of Figure~\ref{fig:cont_view}.  
We computed the average column density, $N_{\rm H_2}^{\rm ave}$ along the filament by averaging the values through the spine data pixels. The physical assumptions used for this calculation are the same as those described in Sect.~\ref{R:HCO_cont}. 
Along the spine of the filament, we constructed an averaged radial profile, from which we determined the FWHM as width, $w_{\rm fil,cont}$, by fitting a Gaussian function. Note that most of $w_{\rm fil,cont}$ shows $\sim$0.1\,pc, but they are marginally resolved at the present spatial resolution of $\sim$0.1\,pc $\times$ 0.09\,pc. $l_{\rm fil,max}$ and $I_{\rm mean,d}$ represent the length of the longest spine and mean 0.87\,mm intensity, respectively. Some of these quantities resulted in continuum-based-line mass, $M_{\rm line}$ (= $\mu m{\rm _H} N_{\rm H_2}^{\rm ave} w_{\rm fil,cont}$), where $m{\rm _H}$ is the atomic hydrogen mass and $\mu$ is the mean molecular weight per hydrogen molecule, 2.7 \citep{Cox2000}. We defined the H$_2$ number density of the filaments as $n_{\rm H_2}$ = $N_{\rm H_2}^{\rm ave}/w_{\rm fil,cont}$. 
It should be noted that quantities derived from a combination of two physical parameters, such as $n_{\rm H_2}$, might carry even higher uncertainties than H$_2$ column densities. While we regard these values as offering insights at an order-of-magnitude discussion, the detections of high critical density HCO$^+$ line provide complementary evidence for the high-density nature of the entire system. Table~\ref{tab:filament} summarizes the physical properties of the identified filamentary clouds.

To compare with the continuum-based physical quantities along the spines of the filaments, we performed velocity characterization with the following procedure in HCO$^{+}$ data. We extracted the HCO$^+$ spectra at local peaks of the 0.87\,mm closest to each Spitzer source position and determined the systemic velocity, $V_{\rm sys}$, as the maximum intensity channel. Because the spectra exhibited predominantly single Gaussian-like structures, this estimation is considered to be robust. Next, we extracted the HCO$^+$ spectra along the spine pixels of the filaments and shifted their central velocities to the above estimated $V_{\rm sys}$. We then derived the average spectrum and performed Gaussian fittings to estimate the velocity width in FWHM, $\Delta v_{\rm FWHM}$. Each target spectra are shown in Figure~\ref{fig:spectra} in the Appendix. Note that deriving the velocity width averaging along the spine pixels on the moment~2 map (see Figure~\ref{fig:MOM2_view}), the difference between the two methods is not significant, within approximately 30\%. Table~\ref{tab:filament} lists the derived velocity information.

The inferred velocity dispersion, $\sigma_{v}$ (=$\Delta v_{\rm FWHM}$/2$\sqrt{2{\rm ln}2}$), is significantly larger than the thermal linewidth, $\sim$ 0.3--0.4\,km\,s$^{-1}$ at a typical molecular cloud temperature of 20--50\,K in the LMC \citep[e.g.,][]{Minamidani_2008}, indicating that the internal gas kinematics are dominated by non-thermal turbulent motion with a velocity-dispersion range of $\sim$1--3\,km\,s$^{-1}$. Neglecting the magnetic field, the virial mass per unit length is given by the equation $M_{\rm line}^{\rm vir}$ = 2$\sigma_{v}^2/G$ \citep[c.f.,][]{Fiege_2000}, where $G$ is the gravitational constant. Our estimated $M_{\rm line}^{\rm vir}$ is within a factor of 2--5 of the continuum-derived $M_{\rm line}$ (Table~\ref{tab:filament}), suggesting a general consistency between the region traced by thermal dust emission and the gas dynamics probed by the HCO$^+$ line. Regarding structures that exceed the line mass of $10^3$\,$M_{\odot}$\,pc$^{-1}$, they are often observed in active star-forming regions in the MW, such as Cygnus-DR21 and W43, referred to as $``$ridge$"$. Such massive filaments are frequently recognized as mini-starburst regions \citep[e.g.,][]{Motte_2018}.

\begin{table}[htbp]
\begin{center}
\caption{Filament parameters }
\label{tab:filament}
\begin{tabular}{lcccccccccc}
\hline \hline
NAME & $l_{\rm fil,max}$ & $w_{\rm fil,cont}$ & $I_{\rm mean, d}$ & $N_{\rm H_2}^{\rm ave}$ & $n_{\rm H_2}$ & $M_{\rm line}$ & $V_{\rm sys}$ & $\Delta v_{\rm FWHM}$ & $M^{\rm line}_{\rm vir}$ & Hub? \\
 & (pc) & (pc) & (mJy\,beam$^{-1}$) & (cm$^{-2}$) & (cm$^{-3}$) & ($M_\odot$\,pc$^{-1}$) & (km\,s$^{-1}$) & (km\,s$^{-1}$) & ($M_\odot$\,pc$^{-1}$) & \\
\hline
Lh10 & 0.90 & 0.25 &  9.04 & 1.4e+24 & 1.8e+06 & 7.7e+03 & 252.0 & 6.6 & 3.7e+03 & \checkmark\\
Lh09 & 1.27 & 0.13 & 15.14 & 2.3e+24 & 5.8e+06 & 6.3e+03 & 234.0 & 6.0 & 3.0e+03 & \checkmark\\
Lh08 & 1.31 & 0.31 &  1.92 & 3.0e+23 & 3.1e+05 & 2.0e+03 & 237.5 & 3.8 & 1.2e+03 & \checkmark\\
Lh07 & 1.48 & 0.16 &  7.46 & 1.1e+24 & 2.3e+06 & 4.0e+03 & 239.5 & 5.5 & 2.6e+03 & \checkmark\\
Lh06 & 2.04 & 0.19 &  4.45 & 6.8e+23 & 1.2e+06 & 2.8e+03 & 238.0 & 3.7 & 1.2e+03 & \checkmark\\
Lh05 & 1.92 &    - &  4.78 & 7.3e+23 &       - &       - & 242.0 & 4.9 & 2.0e+03 & \checkmark\\
Lh04 & 2.46 & 0.10 &  2.09 & 3.2e+23 & 1.0e+06 & 7.3e+02 & 237.5 & 4.1 & 1.4e+03 & \checkmark\\
Lh03 & 1.01 & 0.15 &  2.74 & 4.2e+23 & 8.9e+05 & 1.4e+03 & 265.0 & 5.1 & 2.2e+03 & \checkmark\\
Lh02 & 2.10 & 0.18 &  2.20 & 3.4e+23 & 6.3e+05 & 1.3e+03 & 237.5 & 3.6 & 1.1e+03 & \checkmark\\
Lh01 & 2.03 & 0.18 &  2.20 & 3.4e+23 & 6.1e+05 & 1.4e+03 & 237.0 & 5.4 & 2.5e+03 & \checkmark\\
Lm10 & 1.26 & 0.10 &  3.65 & 5.5e+23 & 1.7e+06 & 1.3e+03 & 239.0 & 4.8 & 1.9e+03 & \checkmark\\
Lm09 & 1.59 & 0.15 &  1.84 & 2.9e+23 & 6.2e+05 & 9.4e+02 & 273.5 & 4.1 & 1.4e+03 & \checkmark\\
Lm08 & 1.23 & 0.13 &  2.44 & 3.9e+23 & 9.4e+05 & 1.1e+03 & 256.0 & 4.4 & 1.6e+03 & \checkmark\\
Lm07 & 0.84 & 0.19 &  1.14 & 1.8e+23 & 3.1e+05 & 7.2e+02 & 244.0 & 2.4 & 4.7e+02 &  -         \\
Lm06 & 1.34 & 0.17 &  1.50 & 2.4e+23 & 4.6e+05 & 8.6e+02 & 221.5 & 3.5 & 1.0e+03 &  -         \\
Lm05 & 0.49 & 0.21 &  1.57 & 2.4e+23 & 3.7e+05 & 1.1e+03 & 234.0 & 3.3 & 9.0e+02 &  -         \\
Lm04 & 2.35 & 0.12 &  1.52 & 2.4e+23 & 6.4e+05 & 6.4e+02 & 282.0 & 3.1 & 8.1e+02 &  -         \\
Lm03 & 1.39 & 0.12 &  0.76 & 1.2e+23 & 3.2e+05 & 3.2e+02 & 297.5 & 2.2 & 4.0e+02 &  -         \\
Lm02 & 0.55 & 0.14 &  1.59 & 2.5e+23 & 5.7e+05 & 7.7e+02 & 229.5 & 2.7 & 6.0e+02 &  -         \\
Lm01 & 0.61 & 0.10 &  1.81 & 2.9e+23 & 9.1e+05 & 6.4e+02 & 275.5 & 3.6 & 1.1e+03 &  -         \\
Ll10 & 0.87 & 0.07 &  1.85 & 2.9e+23 & 1.3e+06 & 4.5e+02 & 249.0 & 3.8 & 1.2e+03 &  -         \\
Ll09 & 0.76 & 0.21 &  0.91 & 1.5e+23 & 2.3e+05 & 6.5e+02 & 283.5 & 2.6 & 5.5e+02 &  -         \\
Ll08 & 0.80 & 0.16 &  1.13 & 1.8e+23 & 3.7e+05 & 6.0e+02 & 234.5 & 3.2 & 8.8e+02 &  -         \\
Ll07 &    - &    - &     - &       - &       - &       - & 267.0 &   - &       - &          -\\
Ll06 &    - &    - &     - &       - &       - &       - & 229.5 &   - &       - &          -\\
Ll05 & 0.60 & 0.08 &  1.39 & 2.2e+23 & 8.6e+05 & 4.0e+02 & 275.5 & 3.9 & 1.3e+03 & \checkmark\\
Ll04 &    - &    - &     - &       - &       - &       - & 270.5 &   - &       - &          -\\
Ll03 & 0.65 & 0.24 &  0.85 & 1.3e+23 & 1.8e+05 & 7.0e+02 & 264.5 & 2.6 & 5.7e+02 &          -\\
Ll02 &    - &    - &     - &       - &       - &       - & 273.5 &   - &       - &          -\\
Ll01 &    - &    - &     - &       - &       - &       - & 287.0 &   - &       - &          -\\
\hline
\end{tabular}\\
\end{center}
\end{table}

Taking into account for the filament identification results, the following criteria were used to classify whether the protostar-associated molecular cloud traced in 0.87\,mm/HCO$^+$ is a hub-filament or not. (1) According to the FilFinder analysis, the spines extend in two or more directions centered on the YSOs in some of the targets. This visually simple criterion categorizes Lh09, Lh06, Lh05, Lh04, Lh03, Lh01 and Lm09 as hub filaments.
(2) We made HCO$^+$ moment~1 maps and overlaid the FilFinder spine (see Figure~\ref{fig:MOM1_view} in the Appendix). 
Based on the velocity maps, we visually searched for additional elongated components without spines whose relative velocity is $\gtrsim$1\,km\,s$^{-1}$ shifted from that of the spine velocity, and then we additionally categorized Lh10, Lh08, Lh07, Lh02, Lm10, Lm08, Ll05 as hub filaments. The setting of a velocity difference of $\gtrsim$1\,km\,s$^{-1}$ might seem arbitrary, but it has astrophysical implications. This velocity difference is sufficiently larger than the velocity resolution and surpasses the turbulent line width on sub-scales expected from the standad size-linewidth relation in molecular clouds \cite[e.g.,][]{Larson_1981,Solomon_1987,Heyer_2001}. Therefore, we interpret this as an indicator of the presence of multiple filamentary components with slightly different velocities.
(3) The FilFinder analysis of Lm05, Lm03, Lm02, Ll10, and Ll08 shows branched spines. However, we did not classify them as hub filaments, either because the spines and their associated 0.87\,mm/HCO$^+$ structures are very small similar to the beam size (Figures~\ref{fig:cont_view}, \ref{fig:HCO_view} and \ref{fig:MOM0_view}), or because they are continuous in velocity space (Figure~\ref{fig:MOM1_view}) can be well recognized as a single filament.

In summary, we define hub filament in this study as follows: they are observed partially or entirely in 0.87\,mm continuum and HCO$^+$ emissions, with intersections of multiple filaments, which may occasionally show slight velocity differences among each other, centered around massive YSOs. As highlighted in Sect.~\ref{R:HCO_cont}, their detection indicates an extremely high intrinsic density of the filaments themselves, on the order of 10$^5$--10$^6$\,cm$^{-3}$. The resulting categorization is listed in the rightmost column of Table~\ref{tab:filament}. Given the nature of the methodology, misclassifications can occur on the weak emission targets, inevitably leading to some degree of subjectivity. However, a robust, general trend is that the dense molecular structures associated with the brighter protostars are complex in both spatial and velocity distribution (Figures~\ref{fig:cont_view}, \ref{fig:HCO_view}, \ref{fig:MOM0_view}, \ref{fig:MOM1_view} and \ref{fig:MOM2_view}), which lends itself to identification as hub filaments.
More detailed discussions of astronomical meanings of the hub filaments are provided in Section~\ref{sec:dis}.

\subsection{Relation between the properties of protostars and associated filaments}\label{R:RbtwPF}

\begin{figure}[htbp]
    \centering
    \includegraphics[width=1.0\columnwidth]{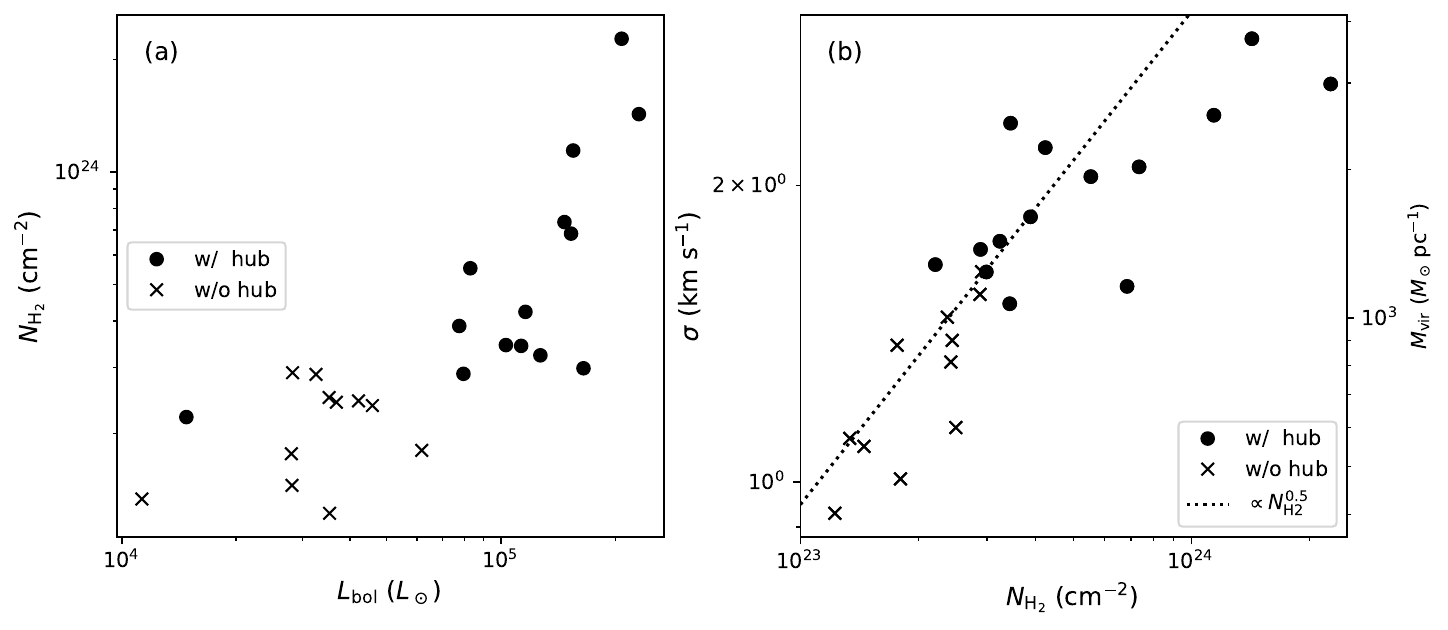}
    \caption{(a) The relation between the bolometric luminosity ($L_{\rm bol}$) of the protostar and the average column density ($N_{\rm H_2}$) along the filament spine. Cross marks and filled circles represent the presence and absence of hub-filaments, respectively. (b) The relation between the column density of the filament and the velocity dispersion derived from HCO$^+$.}
    \label{fig:fil_correlation}
\end{figure}

Figure~\ref{fig:fil_correlation}a illustrates the bolometric luminosity ($L_{\rm bol}$) of massive protostar vs. the average H$_2$ column density ($N_{\rm H_2}$) of the filament crests. Filaments with higher $N_{\rm H_2}$ are generally associated with luminous sources. Further, when we divide our sample into no hub- and hub-filaments, all the identified hub-filaments except Ll05 have luminosities exceeding 5 $\times$10$^4$\,$L_{\odot}$.

Next, we compared the H$_2$ column density with the velocity dispersion (Figiure~\ref{fig:fil_correlation}b) to investigate the molecular gas dynamics. 
It can be inferred that the velocity dispersion increases with increasing H$_2$ column density. Up to a H$_2$ column density of $\sim$5 $\times$10$^{23}$\,cm$^{-2}$, the velocity dispersion follows a power law with an exponent of 0.5. This trend was also reported for objects in a much lower H$_2$ column density regime of 10$^{22}$--10$^{23}$\,cm$^{-2}$ \citep{Arzoumanian_2013} targetting nearby low-mass star-forming region and a few massive sources. We further discuss the H$_2$ column density and velocity dispersion relation in Sect.~\ref{Dis:linewidth_filament}.

\section{Discussion} \label{sec:dis}

Our survey of 30 objects provides valuable insights into high-mass star-forming filamentary molecular cloud systems in the LMC, but the sample bias has to be noted before proceeding to the discussion. As part of the Surveying the Agents of a Galaxy's Evolution (SAGE) Spitzer Legacy program for the LMC (SAGE; \citealt{Meixner_2006}), 1800 unique YSO candidates were identified in the LMC (\citealt{Whitney_2008}; see also \citealt{Gruendl_2009}). \cite{Seale_2009} spectroscopically confirmed 277 YSOs from the sample of 294 YSO candidates with [8.0] $<$ 8\,mag, corresponding to a criterion of high-mass YSOs from \cite{Gruendl_2009}. 
Although our sample only represents 10\% of the total spectroscopically confirmed population, it still gives us a general idea of the overall trend. 
For example, our current criteria identified filamentary structures from which physical quantities could be inferred in 25 out of 30 objects, or 83\% (Table~\ref{tab:filament}). According to the Wilson Score Interval formula, assuming a binomial distribution of the presence or absence of filaments with a sample size of 30, a 95\% confidence interval can be calculated, ranging from 66\% to 92\%.

Most peak positions of dense gas emissions in 0.87\,mm and HCO$^{+}$ objects match the protostellar position, suggesting that we are looking at an early phase of protostar formation widtout significant gas consumption and/or destruction of the parental clouds. 
In addition, a distinct general trend is that more luminous ($\gtrsim$5 $\times$10$^{4}$\,$L_{\odot}$) YSOs are associated with hub filamentary clouds (see Sect.~\ref{R:filament_id}, Table~\ref{tab:filament}, and Figure~\ref{fig:fil_correlation}).  
Similarly, the presence of hub filamentary parental clouds in luminous protostars exceeding a few times 10$^{4}$\,$L_{\odot}$ has been extensively investigated for the Galactic plane objects by \cite{Kumar_2020}. They found hub-filamentary systems with lengths of $\sim$10--20\,pc, total masses of $\sim$10$^{4}$\,$M_{\odot}$ and line masses of $\sim$2 $\times$10$^{3}$\,$M_{\odot}$\,pc$^{-1}$ toward such luminous protostellar objects. Their filament identification is based on Herschel dust continuum measurements that are integrated over the line-of-sight components in both velocity and space, and the spatial resolution did not achieve $\sim$0.1\,pc toward distant ($>$1\,kpc) targets. Therefore, one should carefully compare our results with the Galactic study to provide a hint of the universality of the hub-filament system in the high-mass star formation in the Local Group. 

Our filament identification is superior to the above-mentioned Galactic study if limited to insight into the protostellar-vicinity dense components because the HCO$^{+}$ line selectively captures the inner dense regions with volume densities of 10$^{5}$--10$^{6}$\,cm$^{-3}$ (see Sect.~\ref{R:HCO_cont}) without the contamination of unrelated line-of-sight components as in the Galactic plane. On the other hand, our study cannot distinguish how much total mass or maximum length the filamentary cloud has because the HCO$^{+}$ and 0.87\,mm continuum measurements cannot trace the low-density regime below $\lesssim$10$^4$\,cm$^{-3}$. 
While we acknowledge such limitations in data analysis, it is unnatural for the dense HCO$^{+}$/0.87\,mm structures to exist in isolation; they should be associated with lower-density, larger structures.
For example, early studies in the N159W region, including Lh01, Lh02, and Lh04, showed that the parent cloud of the currently identified hub filament system has a total mass of $\gtrsim$10$^{4}$\,$M_{\odot}$ and a maximum length of $>$10\,pc \citep{Fukui_2015,Tokuda_2019,Tokuda_2022a}. 
Putting together some of the evidence obtained so far from the Galactic Survey and several case studies in the LMC, it appears that massive protostars, whose luminosity is greater than several times 10${^4}$\,$L_{\odot}$, are accompanied by dense hub filaments, which as a cloud form massive systems with a mass of $>$10$^4$\,$M_{\odot}$ and lengths of $>$10\,pc. 
Such luminous protostellar systems have the potential to eventually destroy the parent clouds themselves. The scenario for the formation of hub filament systems and the subsequent evolution of the parent cloud is discussed further in Sects.~\ref{D:GMCevo} and \ref{D:collding_senario}.

\subsection{Linewidth enhancement of massive filamentary clouds}\label{Dis:linewidth_filament}

In this section, we discuss the relation between the velocity linewidth and the H$_2$ column density, $\sigma_{v} \propto N_{\rm H_2}^{0.5}$, as shown in Figure~\ref{fig:fil_correlation}b. 
Although the present beam size is not necessarily high enough to tightly constrain the width ($w_{\rm fil,cont}$), most of the filamentary clouds roughly show a constant width of $\sim$0.1\,pc within a factor of two variations across the targets (see Table~\ref{tab:filament}). 
Considering the radial contraction of an isolated filament, one would expect the width to decrease as the column density increases, but the fact that this is not the case suggests the existence of a mechanism that keeps the filament width constant. 
\cite{Arzoumanian_2013} provided seminal implications based on the velocity structures of filamentary clouds in the solar neighborhood star-forming regions. Their results show a relationship of $\sigma_{v} \propto N_{\rm H_2}^{0.5}$, and argue that the self-gravitating filaments could amplify turbulence due to gravitational contraction and mass accretion from the external environment (see also the case of the Galactic high-mass star-forming region NGC~6334 by \citealt{Arzoumanian_2022}). They focused on the effective Jeans length, determined by the non-thermal (turbulent) motion, which is approximately constant, and discussed why the self-gravitating filaments shrink but the width is close to constant.
Although their results did not include an adequate sample at the higher H$_2$ column density end, our observation of the massive star-forming regions of the LMC allows us to derive remarkably similar results at a higher H$_2$ column density range in many objects. It also allows us to study the dynamics of the high-density (10$^5$--10$^6$\,cm$^{-3}$) gas in the filaments leading to high-mass star formation, which has not been fully explored in either the LMC or MW studies. 

We also observed hub filaments in high-mass systems, suggesting the presence of associated filaments of diffuse gas traced by $^{12}$CO and $^{13}$CO (see also the case of Lh04/N159W-South in \citealt{Tokuda_2019}). Given this, the dense filaments traced in 0.87\,mm and HCO$^+$ and surrounding diffuse gas are likely co-evolving.  
With these results in mind, we discuss the formation of high-mass stars and their parent filament systems in the context of the surrounding gas and star formation conditions in Sections~\ref{D:GMCevo} and \ref{D:collding_senario}.

\subsection{Relation between the filamentary clouds and the evolutionary stages of the parental GMCs}
\label{D:GMCevo}

We examine the relationship between the identified hub-filament systems and the evolutionary stage of their associated molecular clouds to consider the fate of massive star formation and its natal molecular clouds. 
\cite{Kawamura_2009} categorized 272 GMCs in the LMC \citep{Fukui_2008} into three categories by comparing them with H$\;${\sc ii} regions and star clusters: (1) the youngest Type~I, which does not show any signs of high-mass star formation, (2) the intermediate stage Type~II, which only accompanies H$\;${\sc ii} regions, and (3) the most active and final stage Type~III, which is associated with both H$\;${\sc ii} regions and star clusters (see also the original concept proposed by \citealt{Fukui_1999}). The empirically estimated GMC lifetimes for Type~I, II, III are 6\,Myr, 13\,Myr and 7\,Myr, respectively, assuming the number of the three types of GMCs are proportional to the timescale of each stage (\citealt{Kawamura_2009} and see also a recent qualitatively similar result in NGC~628 by \citealt{Demach_2023}). As shown in Table~\ref{tab:target}, the targets in this study are only those that are associated with Type~II and Type~III molecular clouds. Although the classification was estimated based on optical observations, the infrared survey also did not find significant signs of high-mass star formation in the Type~I clouds \citep{Ochsendorf_2016,Ochsendorf_2017}. The number of hub and non-hub filaments associated with Type~II and Type~III is shown in Table~\ref{tab:hubGMC}. 
While systems without hub filaments seem to be evenly distributed between Type~II and Type~III, those with hub filaments appear to be skewed towards Type~III. This tendency suggests that once high-mass star formation initiates somewhere within a GMC, further high-mass star formation can occur at any time without the need for hub filaments; however, exceptionally active star formation, exceeding a luminosity of 5 $\times$10$^4$\,$L_{\odot}$, does not occur until the latest stage of GMC evolution.
Hub-filament systems might act as precursors that eventually disrupt the GMC, impeding its growth.

\begin{table}[htbp]
\begin{center}
\caption{Number of hubs in different evolutionary GMCs}
\label{tab:hubGMC}
\begin{tabular}{lcc}
\hline \hline
             & Type II GMC & Type III GMC \\ \hline
Non-hub      & 7           & 7            \\
Hub filament & 3           & 11           \\ \hline
\end{tabular}
\end{center}
\end{table}

One important caveat in our study is that the original target selection aimed to capture a wide range of infrared luminosities without considering the evolutionary stage of the natal GMCs. A study selecting young stellar objects without bias in each evolutionary stage will be necessary to validate the demographic findings of this work.

\subsection{Hub-filament development at the latest phase of the GMC lifetime}
\label{D:collding_senario}

In line with the above-mentioned evolutionary concept, we consider the conditions under which hub filaments are formed. \cite{Krumholz_2012} simulated the evolution of massive, dense molecular gas systems with a gas mass of $\sim$1,000\,$M_{\odot}$ and a radius of $\sim$0.4\,pc with turbulent perturbations. They demonstrated that supersonic gas flows within the parental cloud produce filamentary structures with a few overdense spots, each of which looks like the turbulent core proposed by \cite{McKeeTan_2003}. 
A complex filamentary system containing massive dense cores can naturally form if a large amount of molecular gas within a tiny volume is available. Because the formed filaments and dense cores are quite dense with a density of $\sim$10$^{5}$--10$^{6}$\,cm$^{-3}$, the evolution time scale follows a free-fall time of the density, $\sim$0.1\,Myr. Because the filaments we identified also have such high densities (see Table~\ref{tab:filament}), it is expected that star formation and gas consumption will proceed rapidly in such a short time determined by the self-gravity \citep[e.g.,][]{Tokuda_2020Tau}, unless they are supported by strong turbulence and/or magnetic fields. Most of the GMC lifetime is dominated by the diffuse molecular gas state at a density of $\sim$10$^2$--10$^{3}$\,cm$^{-3}$, and high-mass star-formation occurs when a high-density region is suddenly formed after a long quiescent phase \cite[e.g.,][]{Kawamura_2009,Chevance_2022}. We thus need to address hypotheses on how to form dense regions with hub-filament morphology during the GMC evolution.

According to the comparison of the CO clouds and the surrounding H$\;${\sc i} gas study \citep{Fukui_2009}, the mass accretion rate onto GMC was estimated to be $\sim$0.05\,$M_{\odot}$\,yr$^{-1}$ assuming a steady state growth owing to the GMC's self-gravity. However, such a constant atomic gas accreting condition is not necessarily realistic \citep[c.f.,][]{Kobayashi_2017,Kondo_2021} because not only the GMC but also H$\;${\sc i} gas envelope have inhomogeneous distribution in nature driven by multi-scale galactic phenomena, such as giant-shell expansions \citep{Kim_2003,Dawson_2013}, and galactic tidal interaction between the LMC and the Small Magellanic Cloud (SMC) \citep[e.g.,][]{Bekki_2007,Fukui_2017,Tsuge_2019}. The thermal instability provides small-scale blobs by the mixing between cold and warm neutral mediums \citep[e.g.,][]{Inoue_2009,Aota_2013,Tokuda_2018}. \cite{Tokuda_2022a} hypothesized that the $``$teardrop inflow$"$ scenario, which is a substructured H$\;${\sc i} gas collision onto GMCs, well explains a simultaneous, numerous cluster formation with hub-filamentary dense clouds in the N159W (Lh04, Lh01, and Lh02) and N159E \citep{Fukui_2019} regions. 

In the typical galactic environment, GMCs are constantly exposed to the risk of gas collisions/flows or interstellar shocks. A developed GMC needs to be supplied with flows to form high-density gas clumps leading to high-mass star formation. Based on ALMA CO observations in the SMC, \cite{Neelamkodan_2021} found a hub-filamentary system with signatures of gas collision. Still, the system only harbors an intermediate stellar system, possibly due to the shortage of the total amount of gas or (column) density. \cite{Enokiya_2021} pointed out that there is a high-H$_2$ column density threshold to initiate active cluster formation (see also recent numerical work by \citealt{Abe_2022}). Structural analysis of CO clouds in the LMC by \cite{Sawada_2018} demonstrated prominent structures such as filaments and clumps during the GMC evolution regardless of the density required for star formation. As disturbances from gas collisions are added to developed clouds, amplified magnetic fileds significantly enhance the effective Jeans mass and then promote the formation of dense hub filaments in a short time of 0.1\,Myr order after the shock compression \citep{Inoue_2018}. 
It is also worth noting that some studies, which have meticulously examined the large-scale gas motions of molecular clouds, have shown that hub filaments are embedded in the regions where interaction between two clouds is ongoing \citep[e.g.,][]{Dewangan_2022,Maity_2022,Maity_2023}. The pre-colliding system possibly corresponds to Stage~I \citep{Kumar_2020}, which comprises more than two systems with individual, simple filamentary clouds in the hub-filament cluster formation evolutionary sequence. 

\section{Summary}\label{sec:summary}
We presented ALMA HCO$^+$~(4--3) and 0.87\,mm continuum data toward 30 massive YSOs with their bolometric luminosities of 10$^4$--10$^{5.5}$\,$L_{\odot}$ in the Large Magellanic Cloud at an angular resolution of $\sim$0\farcs4, corresponding to $\sim$0.1\,pc. Our main results are summarized as follows:

\begin{itemize}
\item[1.] We have confirmed that most of the protostellar sources have dense envelopes with a volume H$_2$ density of $\sim$10$^{5}$--10$^{6}$\,cm$^{-3}$. 
Spatial distributions of HCO$^{+}$~(4--3) are almost the same as those of 0.87\,mm, indicating that the molecular line emission is a useful tracer of dense gas kinematics around high-mass protostellar systems at a 0.1\,pc scale.

\item[2.] The dense gas emission has elongated structures that can be regarded as massive filaments whose line mass is $\sim$10$^{3}$--10$^{4}$\,$M_{\odot}$\,pc$^{-1}$ quantified by the FilFinder algorithm. In particular, filaments with higher H$_2$ column densities tend to have a wider linewidth of HCO$^+$, suggesting an increase in turbulence due to gravitational contraction and gas accretion from the surrounding environment. We found that filamentary systems associated with brighter protostars tend to have more complex hub-like features. The protostellar luminosity threshold for having a hub filamentary system is $\sim$5 $\times$10$^4$\,$L_{\odot}$, consistent with that obtained from a similar Galactic study \citep{Kumar_2020}.

\item[3.]  We have compared the presence or absence of (hub) filament systems with the evolutionary stages of the natal GMCs as revealed by wide-field mapping using a single-dish telescope. The striking finding is that massive hub-filament system are well associated with the latest stage of GMC evolution on their $\sim$30\,Myr lifetimes. 
Based on our preliminary demographics, we suggest that GMCs, grown through various perturbations such as gas collisions, may evolve into systems capable of forming bright, massive star clusters exceeding $\sim$5 $\times$10$^{4}$\,$L_{\odot}$ in their final stages, accompanied by hub-filament precursors.

\end{itemize}

\begin{acknowledgments}
We would like to thank the anonymous referee for useful comments that improved the manuscript. This paper makes use of the following ALMA data: ADS/JAO. ALMA\#2019.1.01770.S. ALMA is a partnership of ESO (representing its member states), the NSF (USA), and NINS (Japan), together with the NRC (Canada), MOST, and ASIAA (Taiwan), and KASI (Republic of Korea), in cooperation with the Republic of Chile. The Joint ALMA Observatory is operated by the ESO, AUI/NRAO, and NAOJ. This work was supported by a NAOJ ALMA Scientific Research grant Nos. 2022-22B, Grants-in-Aid for Scientific Research (KAKENHI) of Japan Society for the Promotion of Science (JSPS; grant No. JP18H05440, JP19K14760, JP20H05645, JP21H00049, JP21H00058, JP21H01145, and JP21K13962). The material is based upon work supported by NASA under award number 80GSFC21M0002 (M.S.).
\end{acknowledgments}

\appendix

\section{HCO$^{+}$ spectra and moment maps}

\begin{figure}[htb!]
    \centering
    \includegraphics[width=1.0\columnwidth]{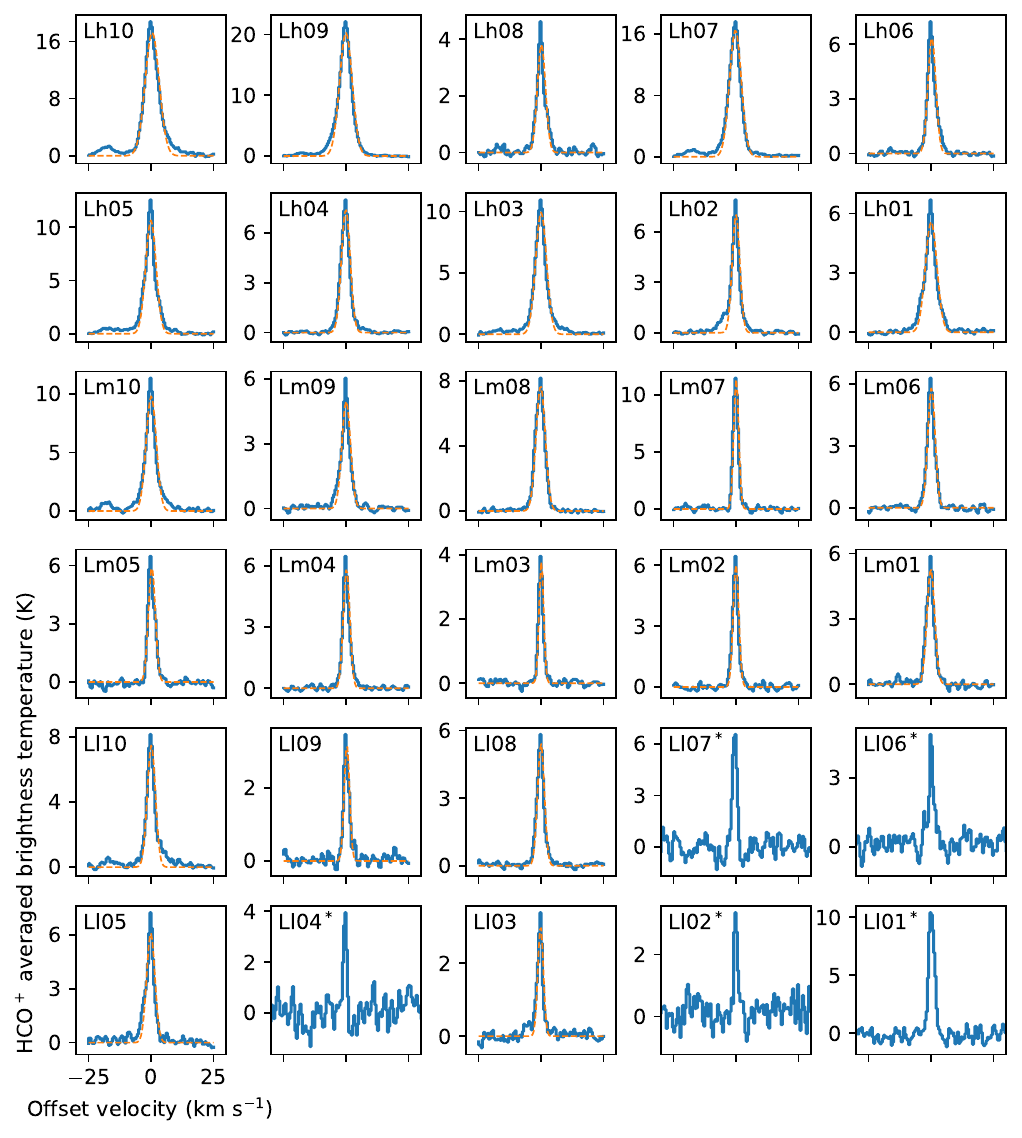}
    \caption{
    Average spectra of HCO$^+$(4--3) along each filament spine are shown in blue solid profile. Orange dashed lines show the Gaussian fitting results. For the targets with the asterisk mark ($*$) just after the name label, we illustrated the HCO$^+$ spectra toward the nearest neighbor of 0.87\,mm peak from the YSO.}
    \label{fig:spectra}
\end{figure}

Figure~\ref{fig:spectra} shows the average HCO$^{+}$~(4--3) spectra along the FilFinder spine (see Sect.~\ref{R:filament_id}). For some bright sources, there are additional blue-shifted components at a relative velocity of $\sim-$15\,km\,s$^{-1}$. This is SO$_2$ emission at the rest frequency of 356.7552\,GHz, instead of an extra velocity component of the HCO$^{+}$. The velocity fitting analysis is unaffected because the SO$_2$ emission is sufficiently weak.

\begin{figure}[htb!]
    \centering
    \includegraphics[width=1.0\columnwidth]{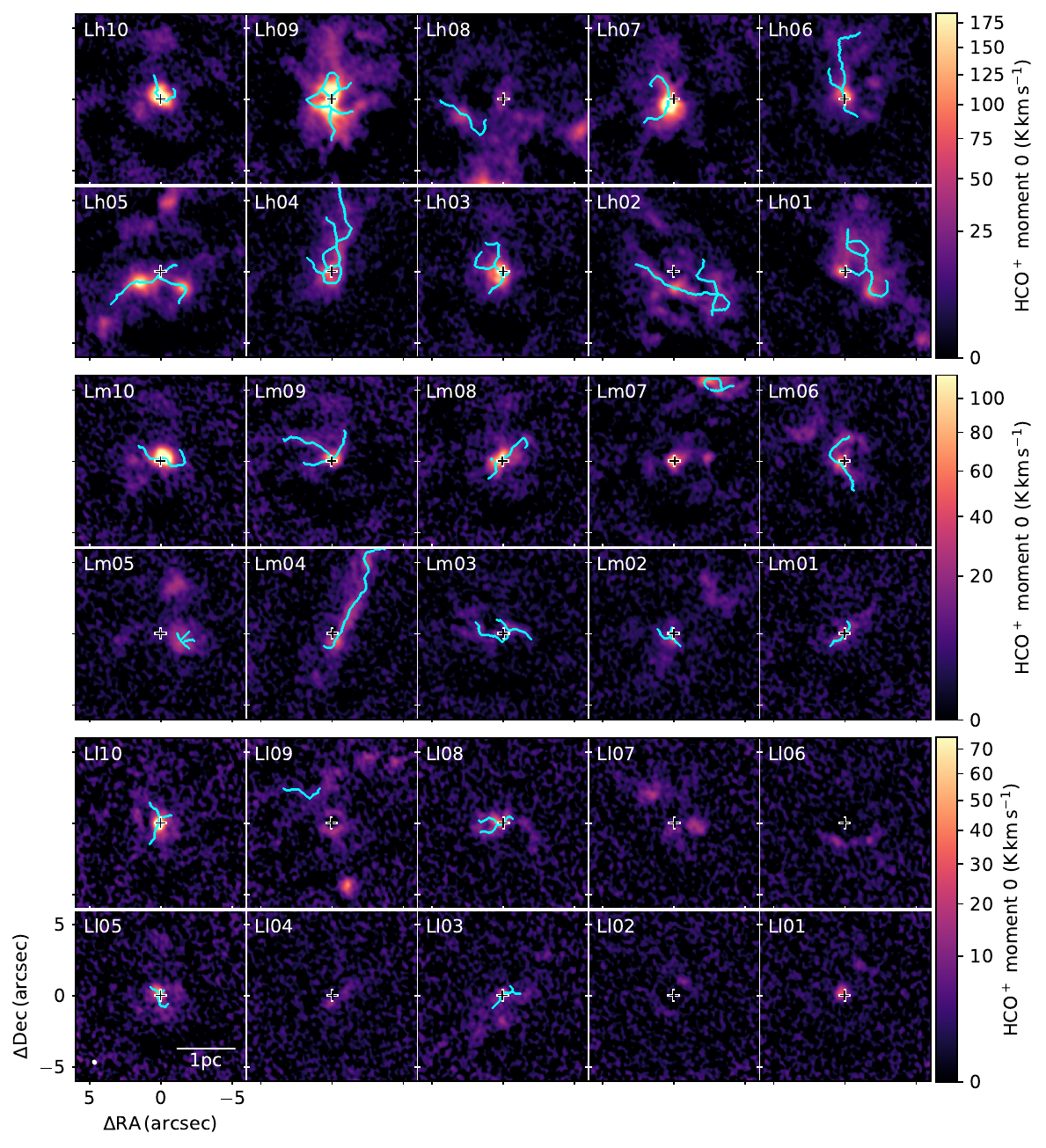}
    \caption{Moment~0 maps of HCO$^+$(4--3) toward high-mass YSOs in the LMC. The integrated velocity ranges are $V_{\rm sys}$ $\pm \Delta v_{\rm FWHM}$ for each source (see Table~\ref{tab:filament}). In the case of $\Delta v_{\rm FWHM}$ not listed objects, namely those for which the filament spine could not be identtfied, a uniform integration range of $V_{\rm sys} \pm 2.5$\,km\,s$^{-1}$ was adopted. Cyan lines show the identified spines of the filament that we used to derive the physical properties (see the text in Sect.~\ref{R:filament_id}).
    }
    \label{fig:MOM0_view}
\end{figure}

\begin{figure}[htb!]
    \centering
    \includegraphics[width=1.0\columnwidth]{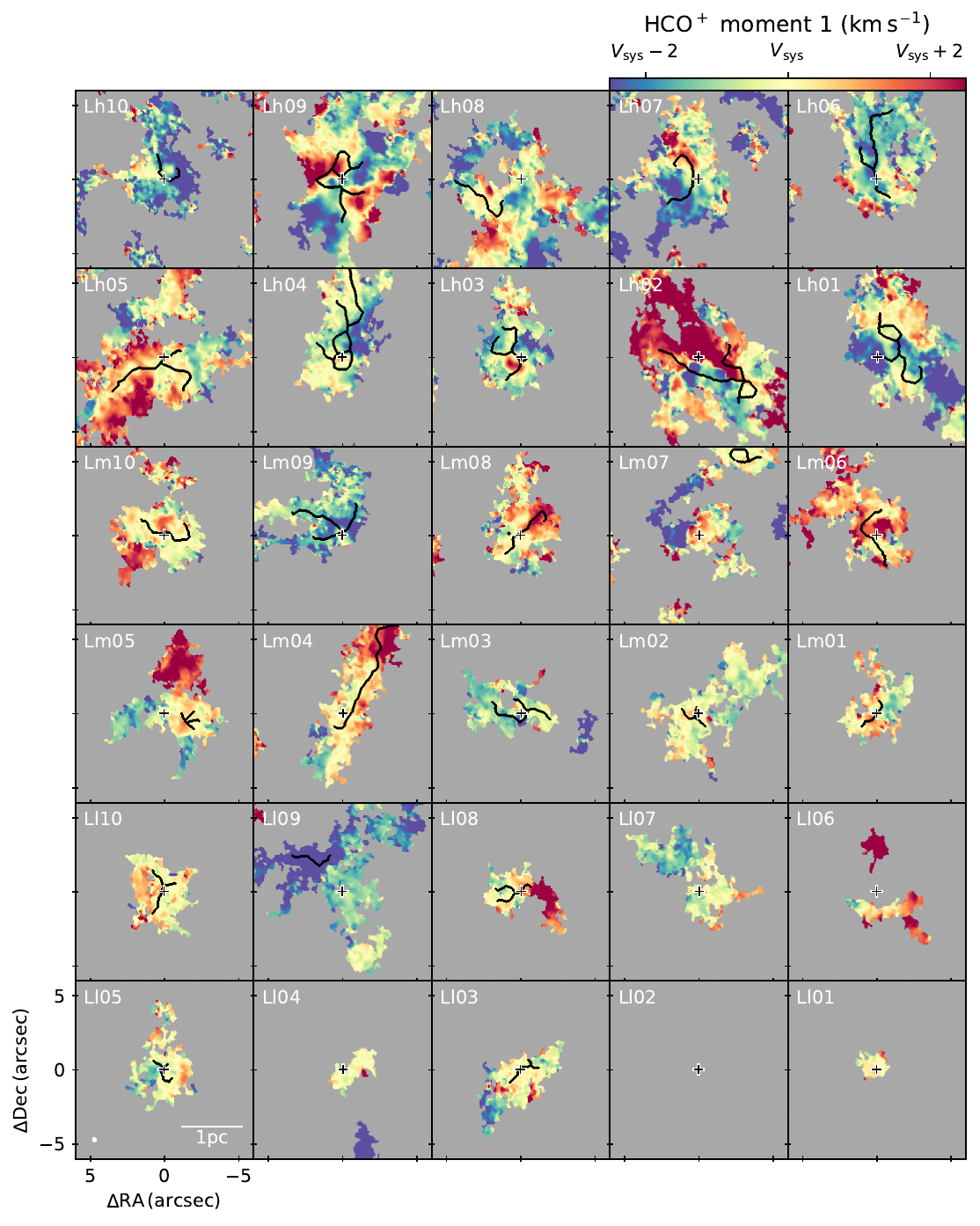}
    \caption{Moment~1 maps of HCO$^+$(4--3) toward high-mass YSOs in the LMC. The angular resolution, $\sim$0\farcs39 $\times$ 0\farcs32, is given by the white ellipse in the lower left corner of the lower left panel. Black crosses represent the position of the YSOs \citep{Seale_2009}. The systemic velocities ($V_{\rm sys}$) for each source are listed in Table~\ref{tab:filament}.　Black lines show the identified spines of the filament that we used to derive the physical properties (see the text in Sect.~\ref{R:filament_id}).}
    \label{fig:MOM1_view}
\end{figure}

\begin{figure}[htb!]
    \centering
    \includegraphics[width=1.0\columnwidth]{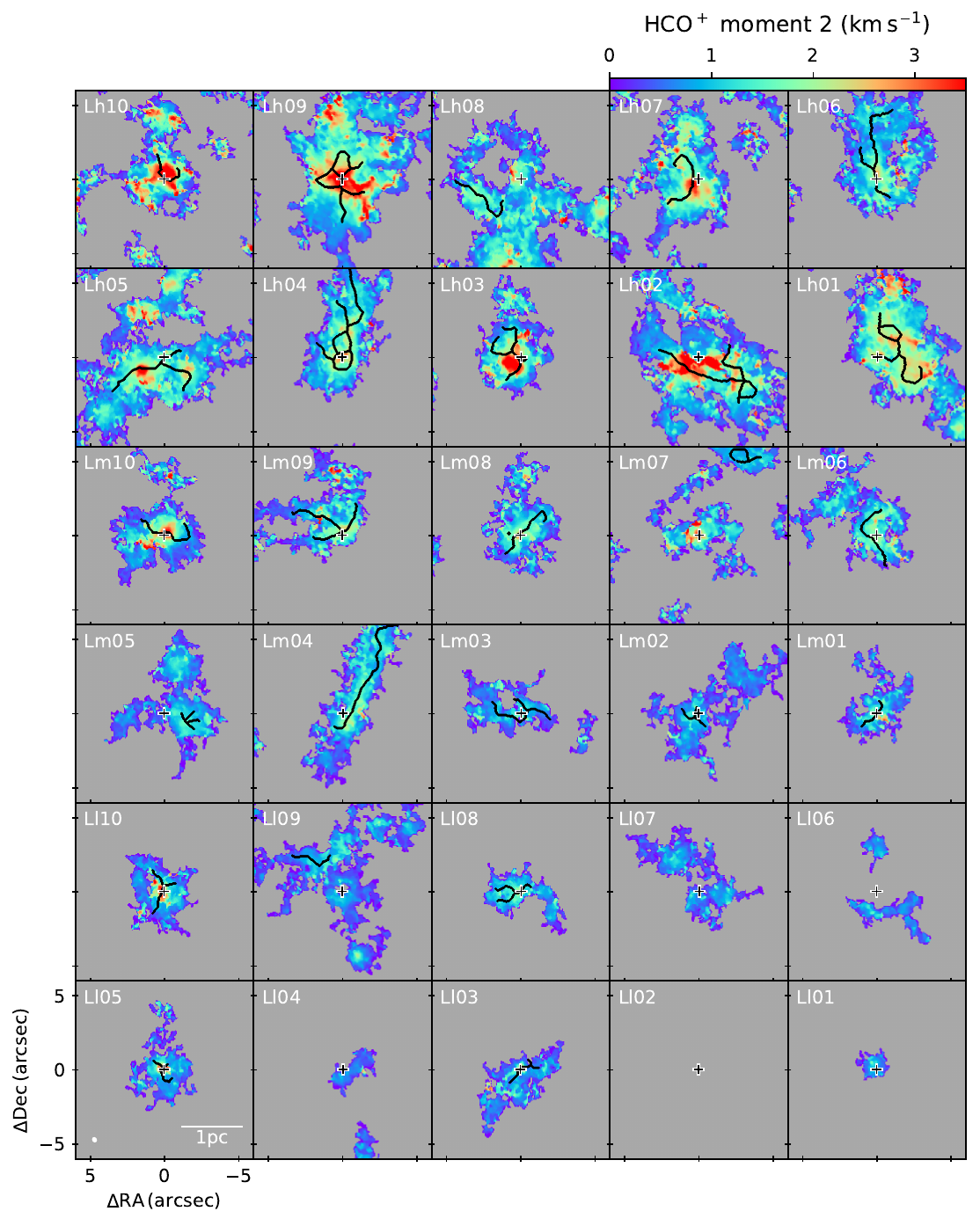}
    \caption{Same as those in Figure~\ref{fig:MOM1_view}, but for moment~2 maps.}
    \label{fig:MOM2_view}
\end{figure}

Figures~\ref{fig:MOM0_view}, \ref{fig:MOM1_view} and \ref{fig:MOM2_view} show the moment~0, 1 and 2 maps, respectively. Typically, at the moment 1 maps, objects located in the upper panels, i.e., those with brighter protostellar luminosities, tend to exhibit complex velocity distributions, often deviating from the systemic velocity by a few km\,s$^{-1}$. Similarly, objects in the upper panels in the moment~2 maps show larger velocity line widths. These trends are consistent with the results in Figure~\ref{fig:fil_correlation}.

%



\software{astropy \citep{Astropy18}}

\bibliography{reference_v01}{}
\bibliographystyle{aasjournal}



\end{document}